%% file: main.tex
\documentclass[twocolumn,journal]{IEEEtran}

\IEEEoverridecommandlockouts                              % This command is only needed if
\usepackage{color}
\usepackage{graphics} % for pdf, bitmapped graphics files
\input{my_sections.tex}

\input{mysymbol.sty}
\usepackage{theorem}
\usepackage{cite}
\usepackage{amsmath}
\usepackage{amssymb}
\usepackage{mathtools}
\usepackage{multirow}
\usepackage{url}
\usepackage{graphics, subfigure, times, amsfonts}
\usepackage{tikz, epic,eepic}
\usetikzlibrary{shapes,arrows}
\usepackage{pgfplots}
\usepackage{color}
\usepackage{hyperref}
\usepackage{epstopdf}
\usepackage{epsfig, amsbsy}
\usepackage{latexsym}
\usepackage{amscd, verbatim}
\usepackage{multirow}
\usepackage{booktabs}

\usepackage{algorithm}
\usepackage{algorithmic}

\newcommand{\QED}{\hfill\ensuremath{\blacksquare}}

\newtheorem{theorem}{Theorem}

\newtheorem{lemma}{Lemma}
\newtheorem{definition}{Definition}

{\itshape}{\rmfamily}
\newtheorem{assumption}{Assumption}
\newtheorem{remark}{Remark}
\newtheorem{condition}{Condition}

\def\forall{\text{for all\ }}

\title{\LARGE A Best-Response Algorithm with Voluntary Communication and Mobility Protocols for Mobile Autonomous Teams Solving the Target Assignment Problem}
\author{Sarper Ayd\i n and Ceyhun Eksin % 
\thanks{S. Aydin and C. Eksin are with the Industrial and Systems Engineering Department, Texas A\&M University, College Station, TX 77843. E-mail:{\tt\small  \; sarper.aydin@tamu.edu; eksinc@tamu.edu}. This work was supported by NSF CCF-2008855.%
}}

\begin{document}
\normalsize
\maketitle

%%%%%%%%%%%%%%%%%%%%%%%%%%%%%%%%%%%%%%%%%%%%%%%%%%%%%%%%%%%%%%%%%%%%%%%%%%%
%%%   A   B   S   T   R   A   C   T   %%%%%%%%%%%%%%%%%%%%%%%%%%%%%%%%%%%%%
%%%%%%%%%%%%%%%%%%%%%%%%%%%%%%%%%%%%%%%%%%%%%%%%%%%%%%%%%%%%%%%%%%%%%%%%%%%
%
\begin{abstract}
We consider a team of mobile autonomous robots with the aim to cover a given set of targets. Each robot aims to select a target to cover and physically reach it by the final time in coordination with other robots given the locations of targets. Robots are unaware of which targets other robots intend to cover. Each robot can control its mobility and who to send information to. We assume communication happens over a wireless channel that is subject to fading and failures. Given the setup, we propose a decentralized algorithm based on decentralized fictitious play in which robots reason about the selections and locations of other robots to decide which target to select, whether to communicate or not, who to communicate with, and where to move. Specifically, the communication actions of the robots are learning-aware, and their mobility actions are sensitive to the success probability of communication. We show that the decentralized algorithm guarantees that robots will cover their targets in finite time. Numerical simulations and experiments using a team of mobile robots confirm the target coverage in finite time and show that mobility control for communication and learning-aware voluntary communication protocols reduce the number of communication attempts in comparison to a benchmark distributed algorithm that relies on communication after every decision epoch.
\end{abstract}

%%%%%%%%%%%%%%%%%%%%%%%%%%%%%%%%%%%%%%%%%%%%%%%%%%%%%%%%%%%%%%%%%%%%%%%%%%%
%%%   E N D     A   B   S   T   R   A   C   T   %%%%%%%%%%%%%%%%%%%%%%%%%%%%%%%%%%%%%
%%%%%%%%%%%%%%%%%%%%%%%%%%%%%%%%%%%%%%%%%%%%%%%%%%%%%%%%%%%%%%%%%%%%%%%%%%%
%

%%%%%%%%%%%%%%%%%%%%%%%%%%%%%%%%%%%%%%%%%%%%%%%%%%%%%%%%%%%%%%%%%%%%%%%%%%%
%%%   S E C T I O N %%%%%%%%%%%%%%%%%%%%%%%%%%%%%%%%%%%%%
%%%%%%%%%%%%%%%%%%%%%%%%%%%%%%%%%%%%%%%%%%%%%%%%%%%%%%%%%%%%%%%%%%%%%%%%%%%
%
\section{Introduction}

\input{introduction.tex}
\begin{figure}
	\centering
	\begin{tabular}{c}
	\includegraphics[width=.9\linewidth]{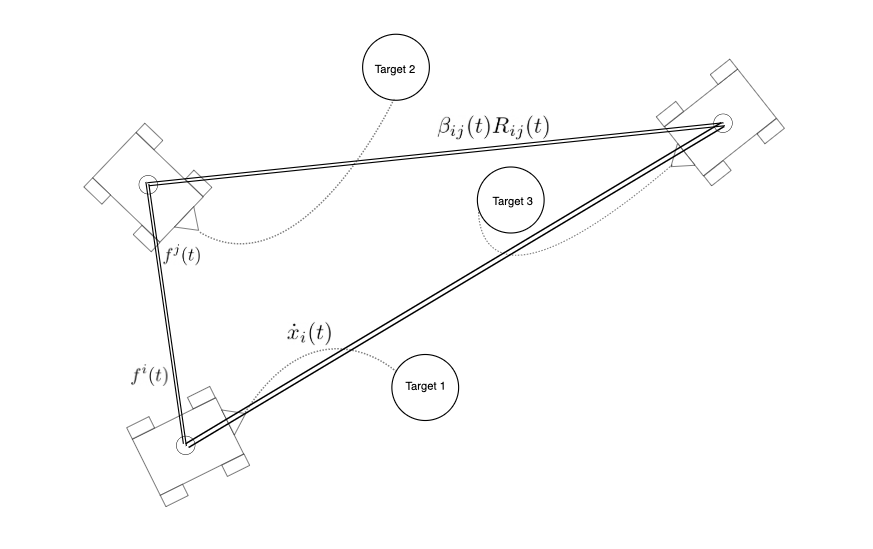} 
	\end{tabular}
	\caption{A team of robots are tasked to solve a target assignment problem. Each robot relies on local estimates of the possible target selection of other robots $f^i(t)$ to select targets. The local estimates are updated based on information received from other robots over a noisy wireless channel subject to failures. The probability of a successful communication from robot $i$ to $j$ ($\beta_{ij}(t)R_{ij}(t)$) depends on their locations ($x_i(t),x_j(t)$), and flow rate ($\beta_{ij}(t)$) determined by robot $i$.}\vspace{-14pt}
	\label{fig_draw}
\end{figure}
%%%%%%%%%%%%%%%%%%%%%%%%%%%%%%%%%%%%%%%%%%%%%%%%%%%%%%%%%%%%%%%%%%%%%%%%%%%
%%%   E N D    F I G U R E %%%%%%%%%%%%%%%%%%%%%%%%%%%%%%%%%%%%%
%%%%%%%%%%%%%%%%%%%%%%%%%%%%%%%%%%%%%%%%%%%%%%%%%%%%%%%%%%%%%%%%%%%%%%%%%%%

%%%%%%%%%%%%%%%%%%%%%%%%%%%%%%%%%%%%%%%%%%%%%%%%%%%%%%%%%%%%%%%%%%%%%%%%%%%
%%%   S E C T I O N %%%%%%%%%%%%%%%%%%%%%%%%%%%%%%%%%%%%%
%%%%%%%%%%%%%%%%%%%%%%%%%%%%%%%%%%%%%%%%%%%%%%%%%%%%%%%%%%%%%%%%%%%%%%%%%%%
%
\section{Target Assignment Problem and a Game Formulation}\label{sec:model}
We consider a team of $N$ robots denoted with $\mathcal{N}=\{1,2,\cdots,N\}$ that move on a 2D surface. There are $N$ targets denoted using $\ccalK:=\{1,2,\cdots,N\}$. The goal of the team is to cover all targets. In order for a robot $i\in \ccalN$ to cover a target $k\in \ccalK$, it has to select that target.  We define the selection variable $a_{ik}\in\{0,1\}$ which is equal to 1 if robot $i$ selects target $k$, and is equal to 0, otherwise. Then the team goal to cover all targets is achieved, when the following equations are satisfied,
\begin{align} \label{eq_assignment}
    \sum_{k \in \mathcal{K}} a_{ik} = 1, \, i \in \mathcal{N}, \text{ and }\quad
    \sum_{i \in \mathcal{N}} a_{ik} = 1, \, k \in \mathcal{K}.
\end{align}
If the conditions above are satisfied there is a one-to-one matching between the robot-target pairs.

\myparagraph{Mobility Dynamics and Constraint} Each robot starts at position $x_i(0)\in\reals^2$ and moves to $x_i(t)\in\reals^2$ with a chosen velocity $\dot x_i(t)\in\reals^2$ for $t\in\ccalT:=\{1,\dots, T_f\}$ where $T_f$ is some final time. Assuming uniform time intervals $\Delta t$, we have the following mobility dynamics,
\begin{equation}\label{eq_mobility}
    x_i(0)+\sum_{s=1}^{t} \dot{x}_i (s) \Delta t= x_i (t), \, (i,t) \in \mathcal{N} \times \mathcal{T}.
\end{equation}
Robots determine their velocities in order to reach their selected targets by the final time, i.e.,
\begin{equation}\label{eq_close}
x_i(T_f) =a_i^Tq := \sum_{k \in \mathcal{K}} a_{ik} q_k, \quad i \in \mathcal{N},
\end{equation}
where  $q_k\in\reals^2$ denotes the target $k$'s static location,  the selection vector of robot $i$ is defined as $a_i:=[a_{i1},\dots,a_{iN}]^T$, and $q$ is the target location matrix that is a concatenation of the locations of all targets. The equality in \eqref{eq_close}, when satisfied, ensures that robots are at their selected targets by time $T_f$. 
\myparagraph{Target selection with minimum effort}
Each robot $i$ has to exert effort to cover target $k$ physically. This effort may depend on distances, energy consumption, or existing preferences among targets. The effort required for each pair of robot $i$ and target $k$ has an associated  positive cost value of $d_{ik} > 0$.  Then the team objective can be written as to minimize total {effort} to cover all targets while satisfying the conditions above,
\begin{align} \label{eq_centralized}
\min_{a_{ik},\dot{x}_i(t)}\quad & \sum_{i \in \mathcal{N}}\sum_{k \in \mathcal{K}} d_{ik}a_{ik} \\
\text{s.t.} \quad & \eqref{eq_assignment}-\eqref{eq_close}. \nonumber
\end{align}

%\red{distances may be changed with weights.}
%where we grouped the target selection and velocity decisions in a team decision variable $\bby:=\{\{a_{ik}\}_{i\in \ccalN,k\in\ccalK},\{\dot{x}_i(t)\}_{i\in \ccalN,t\in \ccalT}\}$ belonging to set $\ccalY$. 

The problem in \eqref{eq_centralized} is easy to solve when robots have complete information, i.e. all robots know $d_{ik}$ for all $i$ and $k$. In such a scenario, each robot can compute the optimal solution to \eqref{eq_centralized} and implement its portion of the optimal selection and mobility dynamics.
In general robots cannot be sure of other robots' {costs} to cover the targets, e.g., when it depends on local energy consumption and distances. This means robots need to solve \eqref{eq_centralized} using their local information. Because robots have different and partial information, robots need to reason about each others' selections to make their own selections. Here, we model the reasoning and decision-making of robots using a decentralized game-theoretic learning framework. We first define the target assignment game and then present the decentralized learning framework. 
%
%%%%%%%%%%%%%%%%%%%%%%%%%%%%%%%%%%%%%%%%%%%%%%%%%%%%%%%
\subsection{Target assignment game}\label{sec_target_game}
In a game, robot $i$, who knows its {cost} associated with covering each target  $\{d_{ik}\}_{k\in\ccalK}$, has to compare among its target options $\ccalK$ without the knowledge of the selection of other robots. Here we use the selection vector $a_i\in \reals^N$ to denote the action of robot $i$. The action of robot $i$ belongs to the space of canonical vectors $\bbe_k$ for $k=1,\dots,N$, i.e., $a_i\in \ccalA:=\{\bbe_1,\dots,\bbe_N\}$. We denote the $k$th element of the action by $a_{ik}$ which is equivalent to the definition of the selection variable in \eqref{eq_assignment}. Given the action space, we represent the utility function of robot $i$ as follows,
\begin{align} \label{eq_utility}
\min_{a_{i}\in\ccalA}\quad & u_i(a_{i},a_{-i})=\sum_{k\in \ccalK} d_{ik}\bar{a}_{-ik} a_{ik},
\end{align}
where $a_{-i}:=\{a_j\}_{j\in -i} \in \ccalA^{N-1}:=\prod_{j\in-i} \ccalA$,  $\bar{a}_{-ik}=\max \{a_{jk}\}_{j \in -i}$, and $-i:=\ccalN\setminus \{i\}$. The definition of $\bar{a}_{-ik}$ implies that if there exists a robot $j \in -i$ that selects target $k$, then $\bar a_{-ik}=1$, and otherwise if none of the robots select $k$, then $\bar a_{-ik}=0$. As per this definition and \eqref{eq_utility}, robot $i$'s cost from selecting target $k$ is $d_{ik}$ if there exists another robot selecting the target. Accordingly, the cost of selecting a target $k$ is zero, if no other agent is selecting that target. 

% \green{\large CE: In this formulation, agent $i$ receives a payoff proportional to $d_{ik}$ if someone else is covering that target. This is incorrect. It should be the other way around. In its current form, the best payoff is 0 and there is not difference between targets!! } \blue{This makes sure that robots receive a payoff of zero, if they select to cover a target that is selected by another agent. Otherwise, if no other robot besides robot $i$ selects target $k$, then robot $i$ receives a payoff of $d_{ik}$.} \green{\large Above explanation in blue is what we should capture with the utility function. I suggest we use the following: 
% \begin{align} \label{eq_utility}
% \max_{a_{i}\in\ccalA}\quad & u_i(a_{i},a_{-i})=\sum_{k\in \ccalK} \frac{1}{d_{ik}}(1-\bar{a}_{-ik}) a_{ik},
% \end{align}
% }

%The term $d_{ik}\bar a_{-ik}$ is a constant from the perspective of robot $i$, since it can only control its selection. 
%
The utility of robot $i$ depends on the actions of all the other robots via the term $\bar a_{-ik}$ in \eqref{eq_utility}. This dependence sets up a target assignment game among the robots defined by the tuple $\Gamma=(\ccalN,\{\mathcal{A},u_i\}_{i\in\ccalN})$.
In the target assignment game, the structure of objective function $u_i(a_{i},a_{-i})$ together with the action space $\ccalA$ assume the role of the coverage constraints in \eqref{eq_assignment}. Once robot $i$ selects its target, it can determine its path toward the chosen action as per \eqref{eq_mobility}-\eqref{eq_close} to satisfy mobility constraints. 

\begin{remark} The utility function in \eqref{eq_utility} ensures that each robot incurs a cost of zero for selecting a target when each target is selected by only a single robot. Utility functions with a similar rationale are considered in  \cite{arslan2007autonomous,yazicioglu2016communication,kalam2010game}. In the utility functions considered in \cite{arslan2007autonomous,kalam2010game} for the target assignment game, the payoff from selecting a target $k$ reduces proportional to the total number of players selecting that target. \cite{yazicioglu2016communication} defines binary-valued agent utilities that are equal to 1 if agents are away by a certain distance
threshold from each other considering the area coverage problem where the goal is to cover the nodes of a graph.  %Here, we propose a utility function that yields a zero payoff from a target selection if that target is selected by more than one agent. If a target is selected by only a single agent, that agent's payoff is proportional 
% as binary values. Their utilities is equal to 1 if they are farther away from each other, with a certain distance threshold so that their design guarantees coverage of different nodes in a graph. 
% For a coverage problem 
% \blue{Utility design should capture the interaction between robots and/or individual ranking among decisions for each robot. Similar to \eqref{eq_utility}, \cite{arslan2007autonomous} proposes to reduce individual utilities by dividing total number of robots selecting the same target. \cite{yazicioglu2016communication} defines individual utilities as binary values. Their utilities is equal to 1 if they are farther away from each other, with a certain distance threshold so that their design guarantees coverage of different nodes in a graph. \cite{kalam2010game} represent utilities proportional to distances to the power of number of robots selecting the same target.}
\end{remark}
%\red{Comparison with other studies may be needed.}
%
% where $\bar{a}_{-ik}=\max_{j \in \mathcal{N}\setminus i} a_{jk} $, if there exists an robot $j \in \mathcal{N} \setminus \{i\}$ that selects target $k$, then $a_{-ik}=1$, and $a_{ik}=0$, if none of them selects it. The term $d_{ik}a_{-ik}$ is a constant from the perspective of robot $i$, since it can only control its selection.
% Formally, a game is defined as by the tuple, $\Gamma=(N,(\mathcal{A}_i,u_i))$, where $i \in N$ is the set of players and, in fact, can be assumed to be equal to the set of robots as explained at the previous section. Then, with the finite action space $\mathcal{A}_i$ of each player $i$, and joint action space of all players $\prod_{i \in N} \mathcal{A}_i$, utility function of player $i$, $u _i:\prod_{i \in N} \mathcal{A}_i \rightarrow \mathbb{R} $ maps joint actions to each one's utility.

%%%%%%%%%%%%%%%%%%%%%%%%%%%%%%%%%%%%%%%%%%%%%%%%%%%%%%%%%%%%%%%%%%%%%%%%%%
\section{Decentralized game-theoretic learning in the target assignment game}\label{sec_DECENTRALIZED}

We assume robots do not have time to coordinate their actions apriori. Thus, they learn to select the optimal (equilibrium) actions in the target assignment game via repeated interaction as they move to reach their current target selections. Robots' interactions are determined by the communication model that is subject to fading and pathloss. In such a setting, if robots only move toward the actions they select, the chance of successful information exchanges may significantly  diminish depending on the target locations. In the following, we propose a decentralized game-theoretic learning algorithm where robots determine who to talk to, and their mobility actions according to the tradeoff between the need to communicate and the goal to reach their selected targets as per \eqref{eq_close}. 

\myparagraph{Decentralized fictitious play (FP) with inertia}  We denote the target selection of robot $i$ at time $t\in \naturals^+$ by $a_i(t)\in \ccalA$. In making its target selection robot $i$ needs to form estimates on the current selection of other robots to evaluate its utility in \eqref{eq_utility}. Similar to the FP algorithm, robot $i$ assumes that other robots act according to a stationary distribution that is determined by their empirical frequency of past actions. We define the empirical frequency of robot $i$ as follows,
\begin{equation} \label{eq_empirical_frequency}
    f_i(t) = (1-\rho_1)f_i(t-1)+\rho_1 a_i(t),
\end{equation}
where $\rho_1\in(0,1)$ is a fading memory constant that measures the importance of current actions. In the centralized FP algorithm, robots best respond, i.e., take the action that minimizes their expectation of their utility computed with respect to the empirical frequencies. However, in a setting with mobile robots, we cannot assume robots have access to the current empirical frequencies of all robots at all times. 

Instead, robot $i$ needs to form estimates of the empirical frequencies based on information received from others. We define the estimate of robot $i$ on robot $j$'s empirical frequency in \eqref{eq_empirical_frequency} as $f^i_j(t)$.  The estimate $f^i_j(t)$ belongs to the space of probability distributions on $\ccalA$ denoted as $\Delta(\ccalA)$. Then the expected utility of robot $i$ with respect to its estimates $f^i_{-i}(t):=\{f^i_j(t)\}_{j\in\ccalN\setminus\{i\}}$ is given by 
\begin{equation} \label{eq_exp_utility}
u_i(a_i, f^i_{-i}(t)) = \sum_{a_{-i}}  u_i(a_i, a_{-i})  f^i_{-i}(t)(a_{-i}).
\end{equation}
In FP with inertia, robots best-respond to the estimated empirical frequencies with some probability $\epsilon_{inertia}\in(0,1)$ $\forall{t} \ge 2$,
\begin{align}\label{eq_response}
a_{i}(t)=
\begin{cases}
\argmin_{a_i\in\ccalA} u_i(a_i,f^i_{-i}(t)) &\;\; \text{w.pr. } 1-\epsilon_{inertia} ,\\
a_{i}(t-1) &\;\; \text{w.pr. } \epsilon_{inertia}.
\end{cases}
\end{align} 
% Robots update their estimates $f^i_{-i}(t)$ based on message exchanges with other robots.
In the following, we describe how robots update their estimates about the empirical frequencies of others $f^i_{-i}(t)$ based on message exchanges with other robots.

\myparagraph{Information exchange and estimate updates}
At each time step $t$, robots update their individual empirical frequency $f_i(t)$ according to  \eqref{eq_empirical_frequency} and let $f^i_{i}(t) = f_i(t)$. After updating its individual empirical frequency, robots attempt to exchange their empirical frequencies with each other. Robot $i$ updates its estimate about robot $j$'s empirical frequency as follows, 
\begin{align}\label{eq_info_ex}
f^{i}_{j}(t)=
\begin{cases}
(1-\rho_2) f^{i}_{j}(t-1)+\rho_2 f^{j}_{j}(t), \, \text{if } c_{ji}(t) =1,\\
f^{i}_{j}(t-1), \, \text{otherwise}, 
\end{cases}
\end{align}
where $\rho_2 \in (0,1)$ is a learning rate, and $c_{ji}(t)$ is a Bernoulli random variable that indicates whether the communication attempt by robot $j$ at time $t$ is successful or not. In this update rule, each robot learns from others and updates its estimate of robot $j$'s selections if robot $j$ is able to transmit its information.
The success probability of a communication attempt at time $t$ depends on the communication protocol and channel statistics that we describe next.

\myparagraph{Learning-aware voluntary communication} Robot $i$ decides on whether it wants to communicate with robot $j$ based on two metrics: novelty of its information  $H_{ii}(t):=||f^{i}_i(t)-a_i(t)||$, and the error that robot $j$ makes in estimating $i$'s empirical frequency $H_{ij}(t):=||f^{i}_i(t)-f^{j}_i(t)||$. In particular, robot $i$ assigns a communication weight $w_{ij}(t)$ to robot $j$ that is equal to zero if novelty and distance conditions are respectively below certain threshold constants $\eta_1\in(0,1)$, and $\eta_2 \in (0,1)$, otherwise the weight is equal to the inverse of the empirical frequency overlap between the two robots defined as $\Delta_{ij}(t) := \max(\delta_1, ||f^{i}_i(t)-f^{i}_j(t)|| )$, where $\delta_1\in(0,1)$ is a positive lower bound on $\Delta_{ij}(t)$.
%\red{In particular, robot $i$ assigns a communication weight $w_{ij}(t)$ to robot $j$ that is equal to zero if both novelty and overlap metrics are above certain threshold constants ($\delta_1 \in (0,1)$, $\delta_1\in(0,1)$), and otherwise is equal to the ratio of novelty and overlap metrics,}
%\blue{In particular, robot $i$ assigns a communication weight $w_{ij}(t)$ to robot $j$ that is equal to zero if novelty metric is above certain threshold constants $\delta_1 \in (0,1)$ and otherwise is equal to the ratio of novelty and overlap metrics,}
%
%\red{\begin{align}\label{eq_w}
%w_{ij}(t)= \begin{cases}
%0, \;\; \text{if $\Delta_{ii}(t) <\delta_1$ and $\Delta_{ij}(t)<\delta_1$,}\\
%\frac{\Delta_{ii}(t)}{\Delta_{ij}(t)},  \;\; \text{otherwise.}
%\end{cases}
%\end{align}}
\begin{align}\label{eq_w}
w_{ij}(t)= \begin{cases}
0, \;\; \text{if $H_{ii}(t) \le \eta_1$ and $H_{ij}(t)\le \eta_2$,}\\
\frac{1}{\Delta_{ij}(t)},  \;\; \text{otherwise.}
\end{cases}
\end{align}
%\red{Sarper: Blue is the actual version in the code. My concern from the beginning, convergence analysis requires the the distance $||f^i_i(t)-f^j_i(t)|| \le  somedelta$ It means each robot should assure its own estimate be close the others' estimate on its own selection. So, the question is whether we should satisfy theoretical requirements or just leave here as in blue by weight update rules as we implement in code.  }
%\green{CE: leave as in code.}
%
% \red{Sarper: In reality, we need a something like $\Delta^{ij}(t)=||f^{i}_i(t)-f^{j}_i(t)||$.  If we add such a condition here, then it contradicts with the next sentence I marked with red. It is possible to put this condition only at  convergence. This part directly affects also the explanation of algorithm. }
We provide an intuition for the above threshold rule. The novelty $H_{ii}(t)$ measures the change in the empirical frequency of robot $i$.  $H_{ii}(t)$ gets smaller when robot $i$ repeatedly selects the same target as per \eqref{eq_empirical_frequency}. Together with the condition that $H_{ij}(t)$ needs to be smaller than $\eta_2$, the above threshold rule checks that if robot $j$ needs further information from $i$ in predicting $i$'s target selection accurately. In the case that these thresholds are not met, i.e., $H_{ii}(t)>\eta_1$ or $H_{ij}(t)>\eta_2$, then the communication weight depends on the overlap metric $\Delta_{ij}(t)$. The overlap metric is the estimated similarity between the empirical frequencies of robots $i$ and $j$. If $\Delta_{ij}(t)$ is small, then two robots are likely to select the same targets according to robot $i$. The smaller $\Delta_{ij}(t)$ is, the more important it is for robot $i$ to coordinate the selection of the targets with robot $j$ so that robots $i$ and $j$ do not select to cover the same target. The constant $\delta_1$ puts a cap on the communication weight that a single robot $j$ can have.

In order to compute the communication weight \eqref{eq_w}, robot $i$ needs access to its own empirical frequency $f_i(t)$, its estimate of $j$'s empirical frequency $f^i_{j}(t)$, and robot $j$'s estimate of $i$'s empirical frequency $f^j_i(t)$. Robot $i$ can locally compute $f_i(t)$ and $f^i_{j}(t)$ using \eqref{eq_empirical_frequency} and \eqref{eq_info_ex}, respectively. For computing $f^j_i(t)$, here we devise an acknowledgement protocol where we assume the receiving robot ($j$) sends an ``ACK'' signal to the sender ($i$) upon successful communication. Given this protocol and the initial estimate of $j$ on $i$'s empirical frequency $f^j_i(0)$, robot $i$ (sender) can keep track of the value of $f^j_i(t)$ by using the update rule in \eqref{eq_info_ex} with indices $i$ and $j$ exchanged. 

% \red{We note that the above computation only uses local estimates stored by robot $i$. Since, each robot $j$ can acknowledge that it has received the information $f^i_i(t)$ sent by robot $i$, and given the apriori information $f^j_i(0)$ and $\rho_2$, then each robot $i$ keep track of current value of estimate $f^j_i(t)$ with storing as a copy of them in memory.} 

At each step, robot $i$ computes a communication weight for all robots as per \eqref{eq_w}. Together these weights $\{w_{ij}(t)\}_{j\in\ccalN\setminus\{i\}}$ determine the relative importance of communicating with other robots. Next we explain how these weights are used in determining flow rates.  

We consider point-to-point communication among robot $i$ and robot $j$ with a rate function $R_{ij}(x_i(t),x_j(t))$ that determines the amount of information robot $i$ can send to robot $j$ at time $t$. Robot $i$ can choose the routing rate $\beta_{ij}(t) \in [0,1]$ that controls the fraction of time robot $i$ spends to send information to robot $j$ at time $t$. 
%We define the Bernoulli random variable $c_{ij}(t) \in \{0,1\}$ that determines whether there is a communication link between robot $i$ and robot $j$ at time $t$. 
The probability of existence of a communication link is given by a Bernoulli random variable, that depends on the rate function and the routing rate,
%\red{Let $G(t)=(\mathcal{N},\ccalE,t)$ be a time-varying communication network whose nodes are robots, and edges are communication links between any robots $(i,j) \in  \mathcal{N} \times \mathcal{N}\setminus\{i\}$} \green{CE: Not sure where you use or need this notation.}.
%\blue{Sarper: I believe the way $c_{ij}(t)$ defined itself requires firstly a definition of network. There may be no need for that as well. } \green{Why? I don't think so.}
%
\begin{equation}\label{eq_channel}
\mathbb{P}(c_{ij}(t)=1)=\beta_{ij}(t)R_{ij}(t) =  \beta_{ij}(t)\, e^{-r ||x_{i}(t)-x_{j}(t)||^2_2}
\end{equation}
where $r>0$ is the channel fading constant. 
 
Robot $i$ allocates its routing rate by solving the following optimization problem
\begin{align}\label{eq_rout}
  \max_{\substack{\beta_{ij}(t) \in [0,1] \\ \sum_{j}\beta_{ij}(t) \le 1}}  \, \prod_{j \in \mathcal{N} \setminus \{i\}} \mathbb{P}(c_{ij}(t)=1),
\end{align}
where the total flow rate is $1$, as the sum of time fractions cannot exceed 1 by the definition of flow rates. 
Given the structure of the communication channel in \eqref{eq_channel}, we incorporate the communication weights $ w_{ij}(t)$ into \eqref{eq_rout}. Then, the optimization problem can be reduced to the following by taking the logarithm of the product term and weighting each term by $w_{ij}(t)$, 
\begin{align} \label{eq_rout_2}
     \max_{\substack{\beta_{ij}(t) \in [0,1] \\ \sum_{j}\beta_{ij}(t) \le 1}}  \,  \sum_{j \in \mathcal{N} \setminus \{i\}} w_{ij} (t)\log(\beta_{ij}(t)).
     \end{align}
Robot $i$ solves the concave optimization problem in \eqref{eq_rout_2} at each time step $t$ to determine the flow rates $\beta_{ij}(t)$. Due to the threshold condition in \eqref{eq_w}, the weights in \eqref{eq_rout}  for some of the robots can be zero, which means robot $i$ eliminates a subset of the robots from the communication all together.

In reducing problem \eqref{eq_rout} to \eqref{eq_rout_2}, we observe that the fading terms in communication $r||x_{i}(t)-x_{j}(t)||^2_2$ disappeared. This means the positions $x_i(t)$ and $x_j(t)$ do not play a role in determining the flow rates; even though, the chance of successful communication between robot $i$ and robot $j$ depends on the positions as per \eqref{eq_channel}. Indeed, robots can adjust their positions based on who they need to send information to. In the following we propose such a mobility scheme that accounts for both the dependence of communication success on distances, and reaching the selected targets.

\myparagraph{Communication-aware mobility}
Each robot wants to move towards the location of the target they selected in the target assignment game. At the same time, as per the communication model in \eqref{eq_channel}, the distance between robots is crucial to successfully send their information. 
Since the exact locations $x_{j}(t)$ is unknown to each robot $i$, robot $i$ can use its estimates $f^i_j(t)$ at time $t$ and target locations $q$ together with \eqref{eq_close} to estimate the final location of robot $j$ as shown below,
%\red{Sarper: note to myself: check again time index. I need to define what q is formally.}
\begin{equation}\label{eq_expj}
    \hat{x}^i_{j}(T_f,t)= f^i_j(t)^T q.
\end{equation}
Given the estimates of the locations of other robots $\{\hat{x}^i_{j}\}_{j\in\ccalN\setminus i}$, robot $i$  selects its next heading direction $x_{i}^{dir} (t)$ to jointly optimize communication success and covering the selected target by solving the following problem,
\begin{align}\label{eq_dir}
    \underset{x_{i}^{dir} (t) \in \mathbb{R}^2 }{\min } \, &\sum_{j \in \mathcal{N} \setminus \{i\}} v_{ij}(t) ||x_i^{dir}(t)-\hat{x}^i_{j}(T_f,t)||^2  \\
    &+|| x_{i}^{dir}(t) - \bar{x}_{i}(T_f,t) ||^2, \nonumber
\end{align}
where $\bar{x}_{i}(T_f,t):=a_i(t)^T q$ denotes the location of the target selected by robot $i$, and $v_{ij}(t)>0$ is the weight that robot $i$ puts on moving closer to robot $j$. The weight $v_{ij}(t)$ is computed using the same threshold condition as the communication weight $w_{ij}(t)$ in \eqref{eq_w} but is computed using updated empirical frequency estimates post communication phase. In other words, $v_{ij}(t)$ is an updated version of $w_{ij}(t)$ computed after the information exchange and belief updates. After determining the direction $x_i^{dir}(t)$, robot $i$'s velocity is given by
%\green{CE: Change the weights notation from $v$ to $v$. Also need to mention that they are computed exactly as $w$s, but post-communication. Continue here by defining the notation you used above and then follow by the explanation below. }
%\red{The term $r||x_{i}(t)-x_{j}(t)||^2_2$ in the model is replaced by $v_{ij}(t)  ||x_i^{dir}(t)-\bar{x}_{j}(T_f,t)||^2$ in order to let robot $i$ approach other robots, by determining its direction $x_{i}^{dir} (t)$ using the weights $v_{ij}(t) >0$ again computed by the routine \eqref{eq_w}.} \green{This is a long sentence that tries to convey too much information in one sentence.}
%Since, the locations $x_j(t+1)$ is not certain to robot $i$, it is remodeled as moving towards other robots' expected final positions \red{$\bar{x}_{j}(T_f,t)$}, given the estimates $f^i_j(t)$. \green{You are introducing notation mid-sentence as if it is already used. Using notation that is not defined before is only allowed in equations because reader will look for the following sentence for definitions. In other places you have to introduce notation first and then use it. }
%\begin{equation}\label{eq_velo}
%    \dot{x}_i(t)=\frac{\alpha(t) x_{i}^{dir}(t)-x_i(t-1)}{\Delta t},
%\end{equation}
\begin{equation}\label{eq_velo}
    \dot{x}_i(t)=\frac{\alpha(t)(x_{i}^{dir}(t)-x_i(t-1))}{\Delta t},
\end{equation}
%\red{\begin{align}\label{eq_pos}
%    x_i(t)&=x_i(t-1)+\dot{x}_i(t)\Delta t \nonumber \\
%    &=x_i(t-1)+\alpha(t)(x_{i}^{dir}(t)-x_i(t-1)).
%\end{align}}
%\green{CE: Would remove this. You can just refer to eq. 2 to say how the position at time $t$ is determined.}
%\red{Sarper: This is the last version (Saturday Jan 18), I implemented in the code. now speed is constant. but we can leave here in general form, I am going to write in experiments.}
%
where $\alpha(t)$ is the speed of robots at time $t$, respecting physical constraints. Thus, robots move utilizing \eqref{eq_velo} to both communicate with others and reach their selected targets.

%\red{In overall, after updating their own action vector $f^i_i$, and then as the result of successful communication, each robot update and learn others estimates. Thus, this is the part where the learning process play role in order to have successful one-to-one communication assignment. }

%\green{CE: Try again.}

\subsection{Algorithm}

Algorithm \ref{suboptimal_alg} summarizes the decentralized fictitious play (DFP) algorithm proposed in the previous section that determines mobility (M) and communication (C) decisions of robot $i$, and thus is referred to as MC-DFP for short.

\begin{algorithm}[H]
   \caption{MC-DFP for Robot $i$}
\label{suboptimal_alg}
\begin{algorithmic}[1]
   \STATE {\bfseries Input:}  Physical locations $x_{i0} \in \mathbb{R}^2 \  \forall i \in \ccalN$;  $q_k  \in \mathbb{R}^2$  $\forall k \in \mathcal{K};$ the parameters $\rho_1,\rho_2, \alpha(t), \eta_1,\eta_2,  \delta_1,  T_f$.
\FOR{$t=1,2,\cdots, T_f$}
  \STATE Select an action $a_i(t)$ using \eqref{eq_response}. 
  \STATE Update $f^{i}_i(t)$ with the selected action via \eqref{eq_empirical_frequency}.
  \STATE Compute weights $w_{ij}(t)$ \eqref{eq_w} \forall $j\neq i$.
\STATE If $w_{ij}(t) \neq 0 $, each robot $i$ decides the routing variables $\beta_{ij}(t)$ \eqref{eq_rout_2} and transmit its empirical frequency $f_i(t)$ with probability of success $\beta_{ij}(t)R_{ij}(t)$.
\STATE Update $\{f^{i}_j(t)\}_{j\in\ccalN}$ using  \eqref{eq_info_ex}.
\STATE Compute the weights $v_{ij}(t)$ using \eqref{eq_w}.
\STATE Determine direction \eqref{eq_dir} and move according to \eqref{eq_velo}.
  \ENDFOR 
   \end{algorithmic}
\end{algorithm}

% \begin{algorithm}[MC-DFP]\label{suboptimal_alg}\hfill\\
% {\bf Initialize:}  Physical locations $x_{i0} \in \mathbb{R}^2 \  \forall i \in \ccalN$;  $q_k  \in \mathbb{R}^2$  $\forall k \in \mathcal{K};$ the parameters $\rho_1,\rho_2, \alpha(t), \eta_1,\eta_2,  \delta_1,  T_f$. \\
% %\indent\indent\indent\indent $\ccalG(0) = (\ccalN,\ccalE_B)$ \\ 
% %\indent\indent\indent\indent  \\
% \indent\indent\indent\indent  \\
% %\indent\indent\indent\indent  \\
% {\bf for} $t=1,2,\cdots, T_f$\\
% 	\begin{enumerate}
% 		\item robot $i$ selects an action $a_i(t)$ using \eqref{eq_response}. 
%     	\item robot $i$ updates $f^{i}_i(t)$ with the selected action using \eqref{eq_empirical_frequency}.
%     	\item robot $i$ determines the weights $w_{ij}(t)$ \eqref{eq_w} for all $j\neq i$.
% 		\item If $w_{ij}(t) \neq 0 $, each robot $i$ decides the routing variables $\beta_{ij}(t)$ \eqref{eq_rout} and transmit its empirical frequency $f_i(t)$ with probability of success $\beta_{ij}(t)R_{ij}(t)$.
% 		\item robot $i$ updates $\{f^{i}_j(t)\}_{j\in\ccalN}$ using  \eqref{eq_info_ex}.
% 		\item robot $i$ determines the weights $v_{ij}(t)$ using \eqref{eq_w}.
% 		\item robot $i$ decides its direction \eqref{eq_dir} and moves according to \eqref{eq_velo}.
% 	\end{enumerate}

% \end{algorithm}

Robots start the updates at each time step with the selection of a target in step 3. In steps 4 and 5, robots determine their current empirical frequencies and their communication weights, which they use to determine their flow rates. In step 6, all robots engage in a round of communication with the determined flow rates. After robots receive new information, they update their estimates about the empirical frequencies in step 7. The updated frequencies are used to determine where robots move next in steps 8 and 9.

MC-DFP has two mechanisms, namely, learning-aware voluntary communication (Steps 5-6) and communication-aware mobility (Steps 8-9), that makes it distinct from prior {decentralized} approaches in team of mobile robots \cite{stephan2017concurrent, kantaros2016distributed, fink2011robust}. In contrast to prior approaches that focus on ensuring probabilistic connectivity for all time steps, the proposed communication and mobility mechanisms make learning of others' selections the goal in MC-DFP. Moreover, MC-DFP algorithm considers realistic communication and mobility models compared to prior decentralized game-theoretic learning schemes, e.g., \cite{eksin2017distributed, swenson2018distributed, otte2020auctions}. 
\begin{remark}
MC-DFP algorithm aims to reduce communication attempts by having agents reason about the value of their local information. In this aspect, the protocols are similar to the voluntary communication protocols considered in \cite{aydin2021decentralized} for DFP. The protocols in this paper are tailored to the target assignment problem solved by a team of mobile agents. In particular, we use mobility protocols to aid communication, and thus learning.   
\end{remark}

\section{Convergence Analysis} \label{sec_con}

% MC-DFP (Algorithm \ref{suboptimal_alg}) involves decentralized mechanisms that determine whom to send information
% to, how to reason about other robots' behavior based on information received, and how to select targets in the target assignment game. 
In the following, we show that MC-DFP (Algorithm \ref{suboptimal_alg}) converges to a rational action profile implying that the constraints of the target assignment problem in \eqref{eq_assignment} and \eqref{eq_close} are satisfied.

We begin by introducing game theoretic concepts that will be used in the convergence analysis. 

\subsection{Game theory preliminaries}

A mixed strategy of robot $i$, denoted with $\sigma_i$, is a probability distribution over the action space, i.e., $\sigma_i \in \Delta(\ccalA)$. The set of joint mixed strategies is given by $\Delta^N(\ccalA) = \prod_{i=1}^N \Delta(\ccalA)$ where we assume the individual strategies are independent. A Nash equilibrium (NE) of the game $\Gamma=(\ccalN,\{\mathcal{A},u_i\}_{i\in\ccalN})$ is a strategy profile such that no individual has a unilateral profitable deviation. 
\begin{definition} [Nash Equilibrium] \label{def_NE}
The joint strategy profile  $\sigma^*=(\sigma_i^*,\sigma_{-i}^*) \in \Delta^N(\mathcal{A})$ is a Nash equilibrium of the game $\Gamma$ if and only if for all $i\in\ccalN$
\begin{equation}
    u_i(\sigma^*_i,\sigma^*_{-i})\le u_i(\sigma_i,\sigma_{-i}^*), \quad \forall \sigma_i \in \Delta(\mathcal{A}).
\end{equation}

A NE strategy profile $\sigma^*$ is defined as pure NE if $\sigma^*=(\sigma_i^*,\sigma_{-i}^*) \in \Delta^N(\ccalA)$, as a probability distribution, gives weight 1 on an action profile $a=(a_i,a_{-i}) \in \ccalA^N$. 
\end{definition}

% A game $\Gamma$ is a potential game if there exists a potential function $u:\prod_{i\in\ccalN}\ccalA\to \reals$, such that the following holds for any pair of actions  $a_i\in\ccalA$ and $a_i'\in\ccalA$ and $a_{-i}\in\prod_{j\in\ccalN\setminus\{i\}}\ccalA$, 
% %
% \begin{equation} \label{eq_potential}
%     u_i(a_i,a_{-i})-u_i(a_i',a_{-i})=u(a_i,a_{-i})-u(a_i',a_{-i}).
% \end{equation}
% %
% \red{Sarper: I believe we cannot claim there exists a potential function, due to the existence of different values of $d_{ik}$. Rather, using best-response potential function is more appropriate, and satisfies weakly acyclic property. I provide a definition for best-response potential game.}

A game $\Gamma$ is a  best-response potential game \cite{voorneveld2000best} if there exists a best-response potential function $u:\prod_{i\in\ccalN}\ccalA\to \reals$, such that the following holds for any actions $a_i\in\ccalA$ and $a_{-i}\in\ccalA^{N-1}:=\prod_{i\in\ccalN\setminus\{i\}}\ccalA$, 

\begin{equation} \label{eq_brpotential}
     \argmin_{a_i\in\ccalA} u_i(a_i,a_{-i})=\argmin_{a_i\in\ccalA} u(a_i,a_{-i}).
\end{equation} 
A best-response potential game with finite set of actions $\ccalA$ is weakly acyclic, i.e., a pure-strategy NE exists, and starting from any $a\in\ccalA^N$, there exists a finite best-response path to a pure-strategy NE. 

We will use a result from \cite{aydin2021decentralized} to show the convergence of MC-DFP. The result in \cite{aydin2021decentralized} hinges on the following condition that an information exchange and belief update protocol needs to satisfy.

\begin{condition}[Condition 1, \cite{aydin2021decentralized}]\label{cond_prob}
There exists a positive probability $\hat{\epsilon}>0$ and a finite time $\hat T$ such that if an agent $j\in\ccalN$ repeats the same action for at least $T>\hat T$ times starting from time $t>0$, i.e., $a_{j}(s)=\bbe_k$ for  $s=t,t+1,\cdots,{t+T-1}$ and $\bbe_k\in\ccalA$, agent $i\in\ccalN$ learns agent $j$'s action with positive  probability $\hat{\epsilon}>0$, i.e., $\mathbb{P} (||a_j(t+T)-f^i_j(t+T)|| \le {\xi}| \ccalH(t)) \ge \hat{\epsilon}$ for any $\xi>0$, where  $\{\ccalH(t)\}_{t \ge 0}$ is a sub-sigma algebra of the Borel sigma algebra $\ccalB$ created by the set $(\ccalA^N \times \ccalG)^t$ of actions and the space of all possible networks $\ccalG$.
\end{condition}

Condition \ref{cond_prob}, if satisfied by a communication and belief update protocol, e.g., in MC-DFP, implies that robot $i$ is able to learn the action of another robot $j$ with high probability when robot $j$ repeats the same action. Condition \ref{cond_prob} requires a communication and belief update protocol to be able to learn given an environment that is static for a finite time horizon. The following result, given in \cite{aydin2021decentralized}, states that any FP-type algorithm with inertia will converge to a pure NE of any weakly acyclic game given that it satisfies this condition. 

% ensures that agents may capture information accurately about other agents' actions before taking an action. Any FP algorithm satisfying this condition in addition to the following ones, converges to pure NE almost surely.

\begin{theorem}[Theorem 1, \cite{aydin2021decentralized}]\label{thm_main}
Let $\{a(t)=(a_1(t),(a_2(t),\cdots,a_N(t))\}_{t\ge1}$ be a sequence of actions generated by a FP-type algorithm with inertia where agents best respond as per \eqref{eq_response} and update local empirical frequencies as per \eqref{eq_empirical_frequency}. Suppose the local empirical frequency estimates of agent $i\in\ccalN$ ($f^i_{-i}(t)$) satisfy Condition \ref{cond_prob}. 
% Suppose the followings are true,
% \begin{itemize}
%     \item The game $\Gamma=(\ccalN,\{\mathcal{A},u_i\}_{i\in\ccalN})$ is weakly acyclic.
%     \item Robots are not indifferent between any pure NE $a^*\in \ccalA^N$, $ \{a^*_i\} = \argmin_{a_i\in\ccalA} u_i(a_i,a^*_{-i})$.
%     \item Condition \ref{cond_prob} hold.
% \end{itemize}
 If the game $\Gamma=(\ccalN,\{\mathcal{A},u_i\}_{i\in\ccalN})$ is weakly acyclic, and agents are not indifferent between any two actions at a pure NE, i.e., the set of minimizers in \eqref{eq_response} is a singleton $ \{a^*_i\} = \argmin_{a_i\in\ccalA} u_i(a_i,a^*_{-i})$, then the action sequence $\{a(t)\}_{t\ge1}$ converges to a pure NE $a^*$ of the game $\Gamma$, almost surely. 
\end{theorem}

\subsection{Convergence}
We establish the convergence of MC-DFP via Theorem \ref{thm_main} by showing that its conditions are satisfied. We begin by showing that the target assignment game is a weakly-acyclic game.
% \blue{We are going to show that target assignment game and MC-DFP hold the given conditions. Firstly, we start by proving the existence of a path to pure NE. }

%\red{{\large CE: Begin lemma environment'i proof'u kapsamayacak. Boylelikle proof italik olmaz. the fact that action profile $a(t)$ stays forever at a pure NE (one-to-one assignment) once reached, and there is a positive probability to reach a pure NE from any action profile $a(t)$ }}
\begin{lemma}\label{lemma_pot}
Target assignment game defined by the tuple $\Gamma=(\ccalN, \{\ccalA,u_i\}_{i\in\ccalN})$ with utility function defined in \eqref{eq_utility} is a best-response potential game. 
\end{lemma}
\begin{myproof} Consider the best-response potential function $u(a_i,a_{-i})=\sum_{k\in \ccalK} \bar{a}_{-ik} a_{ik}$, where $\bar{a}_{-ik}$ and ${a_{ik}}$ are defined as per \eqref{eq_utility}. Since the cardinality of the set of targets and robots are the same, $ |\ccalN|=|\ccalK|=N$, there is always at least one target $k$ uncovered given $a_{-i}\in \ccalA^{N-1}$. Suppose robot $i$ selects to cover one of uncovered targets $\bar{k} \in \ccalK$, so that $a_{i}=\bbe_{\bar{k}}$. Then, it holds that  $u(a_i,a_{-i})=u_i(a_i,a_{-i})=0$ and $\min_{a_i \in \ccalA} u(a_i,a_{-i})= \min_{a_i \in \ccalA} u_i(a_i,a_{-i})=0$. Thus, it satisfies the condition in \eqref{eq_brpotential}.
% $\bbe_{\bar{k}}=\argmin_{a_i\in\ccalA} u_i(a_i,a_{-i})=\argmin_{a_i\in\ccalA} u(a_i,a_{-i})$.
\end{myproof}

By Lemma \ref{lemma_pot}, the game is weakly acyclic so that the existence of a pure NE and a finite best-response path to NE is assured. Next, we show the minimizer in a best-response is a singleton at a pure NE. %We make the following assumptions to show convergence to a pure NE of the game. 

% \red{\begin{assumption}\label{as_single}
% For any pair of actions $a_i\in\ccalA$ and $a_i'\in\ccalA$ and for any $a_{-i} \in \mathcal{A}_{-i}$, it holds that,
% \begin{align}
%   u_i(a_i,a_{-i}) \neq u_i(a'_i,a_{-i}) \;\; \forall i\in \ccalN. 
% \end{align}
% \end{assumption}
% This assumption implies that robots cannot be indifferent between two pure NE action profiles. } \green{CE: Change this to indifference at NE}
%\green{\large This lemma requires positivity of $d_{ik}$s. I changed above so that $d_{ik} \in \reals^+$. If you are OK, erase this comment.}
\begin{lemma}\label{lem_single}
For any pure NE action profile $a^*\in \ccalA^N$ of the target assignment game $\Gamma$, it holds that,
\begin{equation}
  \{a^*_i\} = \argmin_{a_i\in\ccalA} u_i(a_i,a^*_{-i}),
\end{equation}
that is, the minimizer is a singleton.
\end{lemma}
\begin{myproof}
Follows by the fact that the set of pure NE action profiles of the target assignment game constitute of action profiles that covers all the targets. At a pure NE in which all agents cover a single target, agent $i$ does not have an action profile that it can deviate to which will lead to an equivalent cost of zero given that $d_{ik}>0$ for all $k\in \ccalK$.
\end{myproof}

This lemma implies that no single robot can be indifferent between any two actions if other robots play according to the pure NE action profile.

%\red{To begin with, in addition to Assumption \ref{as_single}, } \green{CE: There is no relation between what you say here and what follows besides the fact that they are both assumptions. }

%\red{We assume the distances between physical locations between any two robots to be bounded.}
Next we show that the communication and belief update protocol of MC-DFP in Algorithm \ref{suboptimal_alg} satisfies Condition \ref{cond_prob}.
We make the following two assumptions on bounded distance between robots and targets and on guaranteed success of acknowledgement signals.
\begin{assumption} \label{as_dist}
There exists a positive real number $D >0$ such that $ ||x_{i}(t)-x_{j}(t)|| \le D, \; \forall (i,j,t) \in \mathcal{N} \times \mathcal{N} \setminus\{j\} \times \mathcal{T} $, and $ ||x_{i}(t)-q_k|| \le D, \; \forall (i,k,t) \in \mathcal{N} \times \mathcal{K}\times \mathcal{T} $.
\end{assumption}
%\red{This is a reasonable assumption, since it is not expected to have unbounded distances between any two robots.} 

\begin{assumption}\label{as_ack}
A receiving robot $j \in \ccalN \setminus \{i\} $ can successfully acknowledge if they received the estimates $f^i_i(t)$ from the sender robot $i$ given $c_{ij}(t)=1$.
\end{assumption}

%\red{The above assumption makes sure that acknowledgements are received by the sender robot. This is a critical assumption for robots to keep track of others' estimates about their empirical frequency. Given the assumption robots can  compute the communication metric $H_{ij}(t)$. Further, usage of acknowledgement procedure does not burden robots' limited resources compared to the situation without voluntary communication scheme. Since, each robot sends their empirical frequencies with the complexity at least $O(|\ccalA_i|)$, while the acknowledgement signal is just $O(1)$. In addition, because of the 1-bit acknowledgement signal is cheap, it is not demanding to make sure the ACK signal is sent without failure.} 
This assumption assures that acknowledgement signals are always received without failures. From practical perspective, this assumption is required for the computation of the metric $H_{ij}(t)$. {From a theoretical perspective, this assumption allows us to lower bound the probability of successful communication as per \eqref{eq_channel}.}
% Each robot sends their empirical frequencies with the complexity at least $O(|\ccalA_i|)=O(N)$. This requires handling of routing rates to send them successfully.
The acknowledgement signal has complexity of $O(1)$ so ensuring its transmission is not costly for the receiving agent.

\begin{lemma}\label{lem_cond}
Suppose Assumptions \ref{as_dist} and \ref{as_ack} hold.
Let $\{a(t)=(a_1(t),(a_2(t),\cdots,a_N(t))\}_{t\ge1}$ be a sequence of actions generated by the MC-DFP (Algorithm \ref{suboptimal_alg}). Then, Condition \ref{cond_prob} is satisfied for any $\xi>0$ given small enough $0\le\eta_1<\xi/2$, and $0\le \eta_2\le \xi/2$ such that
if an agent $j\in\ccalN$ repeats the same action for at least $T>\hat T$ times starting from time $t>0$, i.e., $a_{j}(s)=\bbe_k$ for  $s=t,t+1,\cdots,t+T-1$ and $\bbe_k\in\ccalA$, agent $i\in\ccalN$ learns agent $j$'s action with positive  probability $\hat{\epsilon}>0$, i.e., $\mathbb{P} (||a_j(t+T)-f^i_j(t+T)|| \le {\xi}| {\ccalH(t))} \ge \hat{\epsilon}$.
\end{lemma}
\begin{myproof}
See Appendix.
\end{myproof}

\noindent Next, we state our main convergence result for MC-DFP. 
\begin{theorem}\label{cor_physical}
Suppose Assumption \ref{as_dist}-\ref{as_ack} holds. Then, at some finite time $t$, robots implementing MC-DFP (Algorithm \ref{suboptimal_alg} achieve the team goal \eqref{eq_assignment} and cover targets physically \eqref{eq_close}.
\end{theorem}
% \red{Cite Condition 1 and discuss.}
\begin{myproof}
%The proof mainly depends on \cite{aydin2021decentralized}.
By Lemma \ref{lemma_pot}, there exists a finite best-response path to reach pure NE $a=(a^*_i,a^*_{-i}) \in \ccalA^N$ from any joint action profile $(a_i,a_{-i}) \in \ccalA^N$ in target assignment game defined by the tuple $\Gamma=(\ccalN, \{\ccalA,u_i\}_{i\in\ccalN})$. Lemma \ref{lem_single} assures that agents are not indifferent between any two actions at a pure NE. Lastly, Lemma \ref{lem_cond} satisfies Condition \ref{cond_prob}. Thus, by Theorem \ref{thm_main}, the sequence of joint pure action profiles $\{a_t=(a_{it},a_{-it})\}_{t\ge 1}$ converge to a pure NE $(a^*_i,a^*_{-i})$ almost surely in finite time.

At a pure NE, there is a one-to-one assignment of robots to targets. Then, if robots converge to a NE $a^*$ in finite time $t$, the mobility weights $v_{ij}(t)$ are zero, i.e., $v_{ij}(t)=0$ for all $j\in \ccalN$. Thus, each robot goes in the direction of their selected target ($q_k$), without changing $a^*_i=\bbe_k$. By Assumption \ref{as_dist}, robots arrive at their selected target locations by following the mobility dynamics in \eqref{eq_mobility} satisfying \eqref{eq_close} in some finite time $t$. 
%\green{\large CE: This does not read like a proof. In the first paragraph, you start with saying NE is a one-to-one assignment and end with showing that NE is a one-to-one assignment...}
\end{myproof}

{The result above shows that MC-DFP is guaranteed to reach a feasible solution to \eqref{eq_centralized}. Theorem  \ref{cor_physical} provides convergence guarantees for the MC-DFP algorithm despite the fact that robots can choose to cut-off communication based on local statistics as per \eqref{eq_w} or move toward other robots in order to increase communication as per \eqref{eq_dir}-\eqref{eq_velo}. }
%\blue{Show that proof satisfies the conditions in Automatica version}
%instead of their selected targets in order to increase communication as per \eqref{eq_dir}-\eqref{eq_velo}. 

In the next section, we assess the effects of voluntary communication, and learning-aware mobility dynamics in MC-DFP
in terms of convergence time and number of communication attempts using simulations and experiments. 

%\begin{remark}\label{rem_comm_weights}
%The assumption that communication weights remain positive until convergence is in general need not hold in implementation of MC-DFP unless the threshold constant $\delta_1$ in \eqref{eq_w} is equal to zero. For arbitrarily small $\delta_1>0$, we can make sure that communication weights remain positive for arbitrarily long time. However, this is not enough to show almost sure convergence as the probability of all robots stopping communication prior to convergence remains positive. This creates another absorbing state where all robots continue to take the same action, not necessarily Nash, since all robots stopped communicating. In numerical experiments, we observe that such a scenario is indeed possible when $\delta_1$ is relatively large with respect to the fading memory constants $\rho_1$ and $\rho_2$. Still, we find that MC-DFP is highly likely to converge to the pure NE when $\delta_1$ is relatively small compared to the fading memory constants $\rho_1$ and $\rho_2$---see Section \ref{sec_numeric} for details. 
%\end{remark}

%\green{CE: Need to state this as a corollary with a proof. The corollary should say robots reach their targets, i.e., satisfy equation (3) in finite time. }
%%%%%%%%%%%%%%%%%%%%%%%%%%%%%%%%%%%%%%%%%%%%%%%%%%%%%%%%%%%%%%%%%%%%%%%%%%%
%%%   S E C T I O N %%%%%%%%%%%%%%%%%%%%%%%%%%%%%%%%%%%%%
%%%%%%%%%%%%%%%%%%%%%%%%%%%%%%%%%%%%%%%%%%%%%%%%%%%%%%%%%%%%%%%%%%%%%%%%%%%
\section{Simulation and Experiments} \label{sec_numeric}
  %%%%%%%%%%%%%%%%%%%%%%%%%%%%%%%%%%%%%%%%%%%%%%%%%%%%%%%%%%%%%%%%%%%%%%%%%%%
%%%   B E G I N    F I G U R E %%%%%%%%%%%%%%%%%%%%%%%%%%%%%%%%%%%%%
%%%%%%%%%%%%%%%%%%%%%%%%%%%%%%%%%%%%%%%%%%%%%%%%%%%%%%%%%%%%%%%%%%%%%%%%%%%
\begin{figure*}
	\centering
	\begin{tabular}{ccc}
	\includegraphics[width=.3\linewidth]{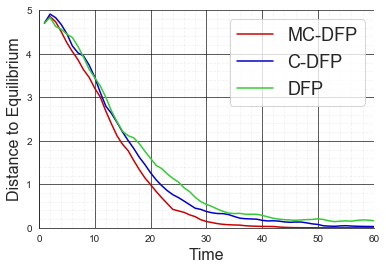} &
	\includegraphics[width=.3\linewidth]{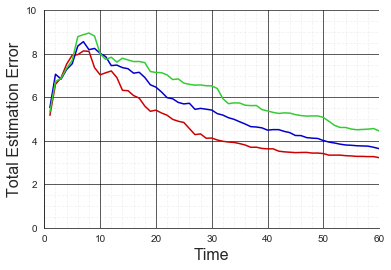} &
	\includegraphics[width=.3\linewidth]{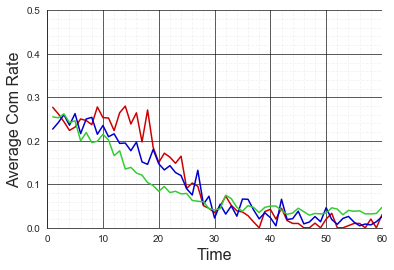}\\
	\includegraphics[width=.3\linewidth]{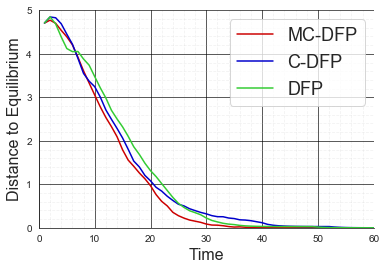}&
	\includegraphics[width=.3\linewidth]{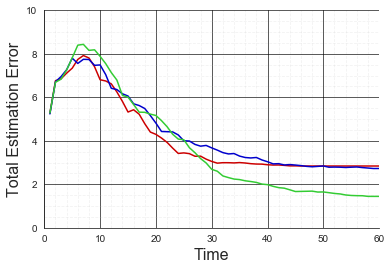}&
	\includegraphics[width=.3\linewidth]{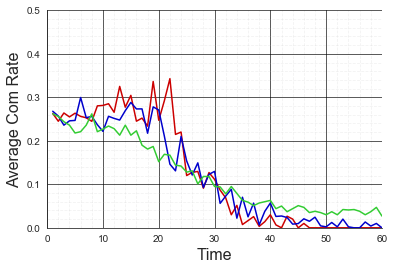}\\
	\end{tabular}
	\caption{Convergence results over 50 replications for Scenario 1 for speeds $\alpha =0.1$ (Top row) and  $\alpha =0.05$ (Bottom row).
	(Left) Convergence of empirical frequencies to pure NE $\sum_{i \in \ccalN} || f^i_i(t)-a^*_i||$.
 (Middle)	Convergence of estimation errors $\sum_{i \in \ccalN} \sum_{j \in \ccalN \setminus \{j\}}||f^i_i(t)-f^j_i(t) ||$.    (Right) Success ratio of communication attempts over time ($\sum_{i \in \ccalN} \sum_{j \in \ccalN \setminus \{j\}} \frac{c_{ij}(t)}{ \bbone_{\beta_{ij}(t) >0}}$). Robots play a pure NE action profile after  $t=40$ on average.} \vspace{0pt}
	\label{fig_conv_sc1}
\end{figure*}
In both simulations and experiments, we assume that the cost of robot $i$ to cover target $k$ is proportional to the distance that it has to traverse, defined as $d_{ik}:=||x_i(0)-q_k||_2^2$.

\subsection{Simulation Setup}

We consider $N=5$ robots and targets, in which robots and targets are positioned according to two different scenarios. In Scenario 1, robots start at the origin $(0,0)$ and targets are located at $(0,1),(1,1,),(1,-1),(-1,1),(-1,-1)$. For Scenario 2, robots are positioned at different starting points $(-0.5,0,),(-0.5,-0.5),(-0.5,0.5),(0.5,0.5),(0.5,-0.5)$, and also targets are given as $(0,0),(-0.5,1.5),(-0.5,-1.5),(0.5,1.5),(0.5,-1.5)$.

The algorithmic parameters $\rho_1$, $\rho_2$, and $\epsilon_{inertia}$ are chosen as $0.4$, $1$, and $0.05$, respectively. The initial empirical frequencies and their estimates ($f^i_i(t)$ and $f^i_{j}(0)\, \forall{(i,j)} \,  \in \mathcal{N} \times \mathcal{N} \setminus \{j\}$) are assigned as uniformly distributed over $5$ targets so that $f^i_i(t)=[0.2, \cdots, 0.2]$ and $f^i_{j}(0)=[0.2, \cdots, 0.2]$. The channel fading constant $r$ is determined as $0.65$. Moreover, each scenario is experimented with different constant speed values over time $\alpha(t)$, that are respectively $0.1$ and $0.05$ for Scenario 1 and $0.05$ and $0.025$ for Scenario 2. Communication threshold constants $(\eta_1,\eta_2)$ are given as $(0.1,0.4)$. We explore the MC-DFP performance with respect to parameters $\rho_1$,  $\rho_2$, $\eta_1$, and $\eta_2$ in Section \ref{sec_parameter}. Lastly, upper bound for $\Delta_{ij}$ in \eqref{eq_w} is selected as $\delta=10$. Targets are assumed to be covered if the Euclidean distance to final positions of robots are within 0.1. 

Given the setup, we compare the performance of MC-DFP algorithm with respect to two decentralized benchmark learning schemes. The first benchmark learning scheme only utilizes learning-aware voluntary communication and does not use communication-aware mobility, i.e., it only moves toward the selected target. We denote this learning algorithm as C-DFP algorithm. The second benchmark algorithm only implements DFP without learning-aware voluntary communication and communication-aware mobility. We denote this learning algorithm as DFP. In DFP, we further replace the voluntary communication protocol in C-DFP by a fixed communication protocol where robot $i$ attempts to communicate at all time steps with equal flow rates for all robots, i.e., $\beta_{ij}(t)=\frac{1}{N-1}=0.25$. 

\subsection{Rate of convergence to an NE and estimation errors}
 Fig.~\ref{fig_conv_sc1}(left) illustrates the convergence to equilibrium in Scenario 1 with top and bottom figures corresponding to speeds 0.1 and 0.05, respectively. All three algorithms converge to a pure NE in all of the 50 cases within the time frame $T_f$.  MC-DFP has a slightly faster average convergence rate. We do not observe a significant effect of robot speed in convergence to NE while it has some effect on communication success as we discuss in the following sections. 
 
 Note that only the benchmark DFP has positive communication weights at all times. This means the total estimation error of robots estimating each others' empirical will go to zero. Bechmark DFP is the only algorithm among the three that guarantees convergence to zero in estimation errors. However, given the communication failures due to fading, diminishing of estimation errors may take a long time to be practically relevant as is evident from the similarity of the estimation errors among the three algorithms in Fig.~\ref{fig_conv_sc1}(Middle). 
 Fig.~\ref{fig_conv_sc1}(Middle) shows the total error robots make in estimating each others' empirical frequencies. Combined with the fact that all learning algorithms converge, i.e., the action profile is a NE, before the final time $T_f$, we can conclude that robots can converge to a pure NE even when there remains gaps between actual and estimated empirical frequencies. That is, the sustained communication attempts in DFP does not provide an advantage over C-DFP and MC-DFP. In summary, DFP comes with unnecessary communication attempts incurring significant energy costs to robots as we explore next.

%%%   B E G I N    F I G U R E %%%%%%%%%%%%%%%%%%%%%%%%%%%%%%%%%%%%%
%%%%%%%%%%%%%%%%%%%%%%%%%%%%%%%%%%%%%%%%%%%%%%%%%%%%%%%%%%%%%%%%%%%%%%%%%%%
\begin{figure}
	\centering
	\begin{tabular}{c}
	\includegraphics[width=.65\linewidth]{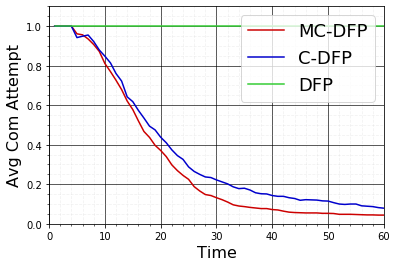} \\
	\includegraphics[width=.65\linewidth]{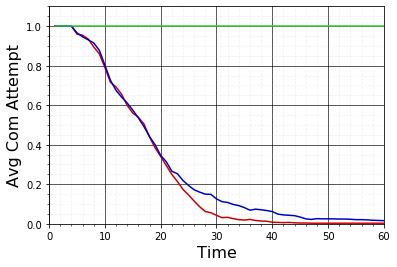} 
	\end{tabular}
	\caption{Average communication attempts per link in Scenario 1 for speeds 0.1 (Top) and 0.05 (Bottom). Average communication attempt per link is obtained by dividing total number of communication attempts at each step by the total number of possible communication attempts, which is equal to 20. The results show average over all 50 runs. MC-DFP reduces the total communication attempts over the entire horizon by a factor of three on average compared to sustained communication in DFP.  } \vspace{0pt}
	\label{fig_com_link}
\end{figure}
\subsection{Effects of learning-aware voluntary communication}
% \red{Sarper: We essentially two different arguments at this section: 1) Even if robots continue to communication, they may not decrease error since their distances increasing that results in failure. 2)Decreasing error after some point is unnecessary, because it is enough  to choose uncovered target at some level of error. }
The total estimation errors with respect to time follow a similar shape for all algorithms---see Fig.~\ref{fig_conv_sc1}(Middle). There is an initial increase in the estimation error starting from an uninformative common prior $f^i_j(0)$ as robots begin to make target selections using best-response with inertia. After reaching a peak around $t=8$, the total estimation error decreases  implying that robots learn the empirical frequencies of  others. In C-DFP and MC-DFP, as robots successfully transmit their empirical frequencies, they begin to reduce communication attempts as per \eqref{eq_w}. Indeed, after time  $t=5$, robots begin to reduce communication attempts in both C-DFP and MC-DFP. By time $t=18$, communication attempt per link drops below 0.5 for both C-DFP and MC-DFP. That is, robots attempt to use a link less than 50\% of the time. The average communication attempt per link shown in Fig.~\ref{fig_com_link} highlights the relative reduction in total cost of communication energy. 

The cease of communication attempts leads to a slow down in descent of total estimation errors in C-DFP and MC-DFP compared to DFP (see Fig.~\ref{fig_conv_sc1}(Middle)). Nevertheless, the slow down does not prohibit convergence to a NE as discussed in the above section. Moreover, when robots are moving faster, we observe that robots have higher total estimation errors in DFP due to fading becoming an important factor early on (compare top and bottom rows of Fig.~\ref{fig_conv_sc1}(Middle)). The intuition for this is as follows. In contrast to DFP, robots allocate communication rates by prioritizing robots based on their need for information in C-DFP and MC-DFP. This helps in obtaining smaller estimation errors faster when fading is important as in the case when robot speeds are fast. 

Fig.~\ref{fig_conv_sc1}(Right) shows the average success ratio of communication attempts with respect to time in the three learning schemes. All learning models start with similar success rates as neither prioritization or mobility has any effect on communication success. Over time, there is a gradual decrease in chance of communication success for all models due to robots moving away from each other toward their selected targets. However, this gradual decrease is faster at the beginning ($t\in(0,20]$) for DFP as robots do not allocate their communication rates by prioritization as they do in C-DFP. After time $t=30$, communication success ratio drops to zero for C-DFP and MC-DFP while DFP retains a small chance of success around 0.05. This is because we let communication success be equal to zero by convention if a communication attempt between  two robots is ceased. 

Overall, the voluntary communication protocol in \eqref{eq_w} saves energy without hampering team performance with appropriately chosen communication threshold constants. 

\begin{figure}
	\centering
	\begin{tabular}{cc}
	\includegraphics[width=.47\linewidth]{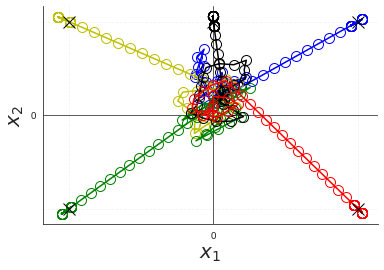} &
	\includegraphics[width=.47\linewidth]{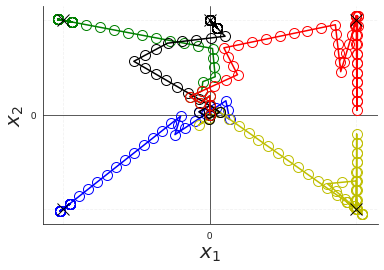} \\
	\includegraphics[width=.47\linewidth]{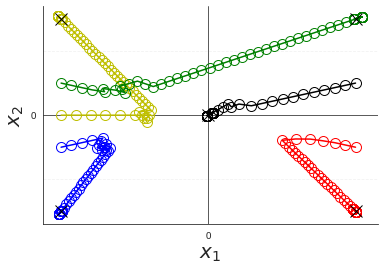} &
	\includegraphics[width=.47\linewidth]{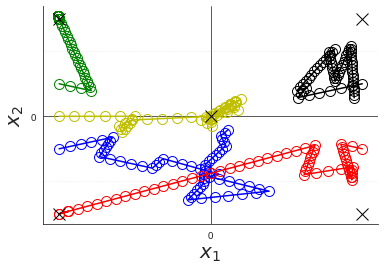}
	\end{tabular}
	\caption{Positions of robots over time in MC-DFP  (Left) and  DFP  (Right) in Scenarios 1 (Top) and 2 (Bottom). In Scenarios 1 (Top) and 2 (Bottom), robots move at speeds $\alpha=0.1$ and $\alpha=0.05$ respectively.  All robots arrive at targets by time $T_f$ in MC-DFP for both scenarios (Left). Targets remain uncovered in DFP for both scenarios (Right). Mobility-aware communication allows quick dissemination of information by evading failures due to pathloss.} \vspace{0pt}
	\label{fig_mov_sc1}
\end{figure}
\subsection{Effects of communication-aware mobility}

Fig.~\ref{fig_conv_sc1}(Right) also demonstrates the effect of mobility on communication success ratio. Specifically, at the beginning $t\in (0,20]$, robots' attempts to overcome fading by moving toward their intended communication targets (receiving robots) yield higher success rate for communication in MC-DFP compared to other algorithms. This high success rate results in lower average communication attempt per link in Fig. \ref{fig_com_link}.

Fig.~\ref{fig_mov_sc1} demonstrates the effects of communication-aware mobility on the team movement for Scenarios 1 and 2. In Scenario 1 (Fig.~\ref{fig_mov_sc1} (Top)), robots start from the same location which means communication failure due to fading is not likely. In Fig.~\ref{fig_mov_sc1} (Top-Left) robots stay close due to the communication-aware direction selections as per \eqref{eq_dir}. In contrast, when robots move toward their selected targets in DFP, we observe robots heading away from each other early on following their best target selections followed by sharp direction changes in Fig.~\ref{fig_mov_sc1} (Top-Right). In Scenario 2 (Bottom), three robots on the left are close to each other but are far from the two robots on the right who are also close to each other. This implies that the robots on the left are highly unlikely to communicate with the  robots on the right at the beginning. In MC-DFP (Bottom-Left), all robots move toward the center target for a long time increasing the chance of successful communication between the initially disconnected robots. This behavior that minds communication highly increases the team's chance to cover each target by final time. In contrast, robot movements are driven by target selections in DFP (Bottom-Right). This reduces the chance of communication between robots on the left with robots on the right, leading to some targets not being covered by final time.

We further analyze the effect of speed on team's likelihood of covering every target in different scenarios. Table \ref{table:coverage} shows that with decreasing speed, convergence is less likely. In particular for Scenario 2 where subsets of robots start distant from each other (high initial fading), likelihood of covering all targets by final time drops for all algorithms. This drop is higher in C-DFP and DFP compared to MC-DFP.

\begin{table}[t]
\centering
\begin{tabular}{@{}l l l l l @{}}\toprule
\multicolumn{1}{ c  }{} & \multicolumn{1}{ c  }{}&     \multicolumn{1}{ c  }{}&            \multicolumn{1}{ c }{Coverage}\\
\cmidrule{3-5}
\multicolumn{1}{ c  }{} & \multicolumn{1}{ c }{Speed} &\multicolumn{1}{ c }{MC-DFP}   &\multicolumn{1}{ c }{C-DFP} &\multicolumn{1}{ c }{DFP} \\
\cmidrule{3-5}
%\multicolumn{1}{ c  }{} & 0.05  & 5  &\\ \midrule
\multicolumn{1}{ c  }{Scenario 1}& 0.1 & 1.00 &0.96 &0.86 \\
\multicolumn{1}{ c  }{}& 0.05 & 0.98 &0.92  &0.90 \\ 
\multicolumn{1}{ c  }{Scenario 2}& 0.05 & 0.96  &0.92 &0.94\\
\multicolumn{1}{ c  }{}& 0.025 & 0.74 &0.58 &0.42 \\
\bottomrule
\vspace{0pt}
\end{tabular}
%ans =
%
%  361.1600  171.8600  220.8600  103.8200
%
%>> mean((convergence_time(:,51:100)'))
%
%ans =
%
%  328.4800  178.7600  212.1400   94.8000
%
\caption{Chance of successful physical coverage by final time}
\label{table:coverage}
\end{table}

\subsection{Parameter Sensitivity} \label{sec_parameter}

\begin{figure}
	\centering
	\includegraphics[width=1\linewidth]{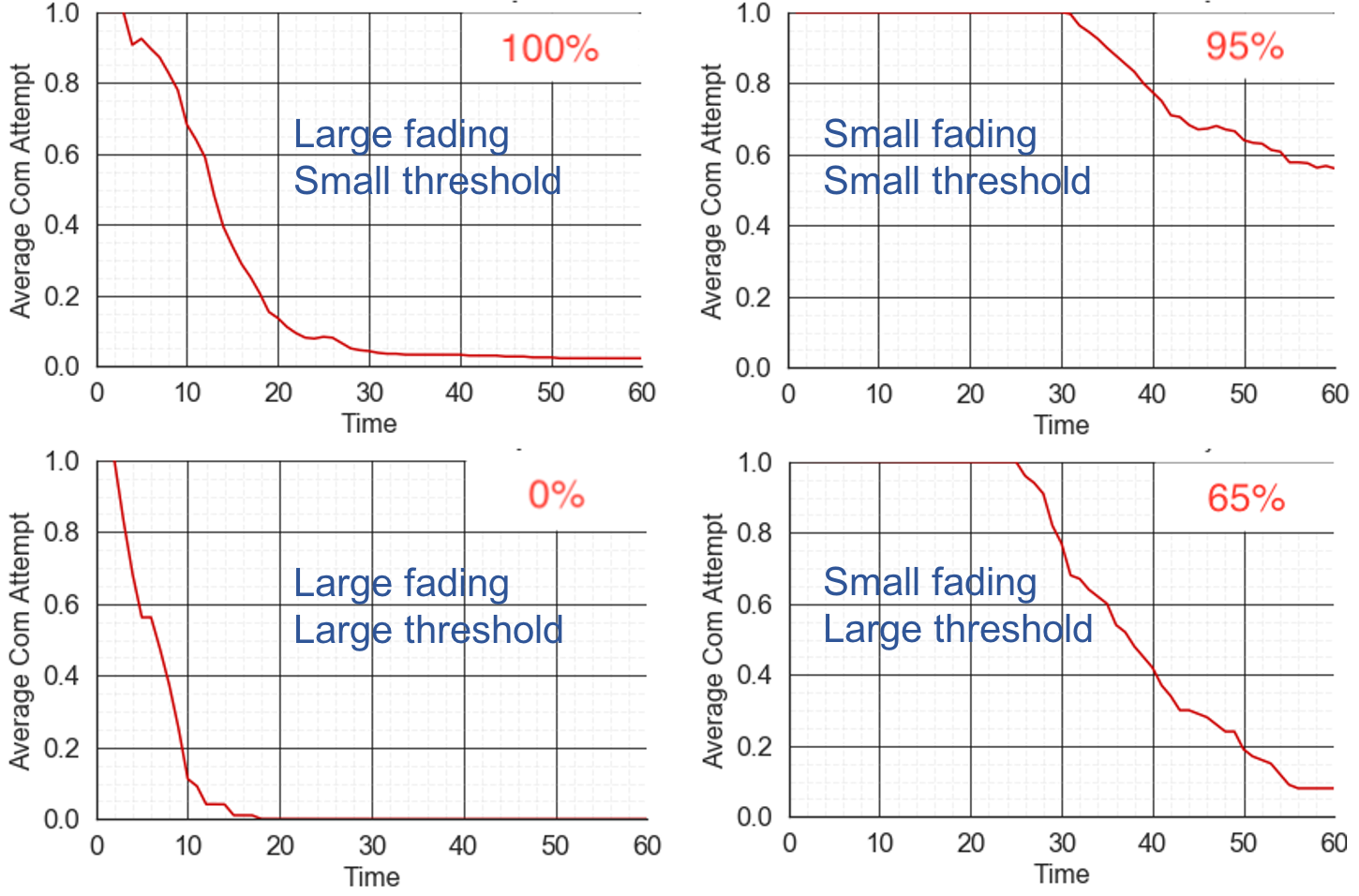}
% 	\begin{tabular}{cc}
% 	\includegraphics[width=.47\linewidth]{dfp_com_aware/figs/11.png} &
% 	\includegraphics[width=.47\linewidth]{dfp_com_aware/figs/12.png} \\
% 	\includegraphics[width=.47\linewidth]{dfp_com_aware/figs/21.png} &
% 	\includegraphics[width=.47\linewidth]{dfp_com_aware/figs/22.png}
% % 	\includegraphics[width=.3\linewidth]{dfp_com_aware/figs/dfp_mov.png}
% 	\end{tabular}
	\caption{Average communication attempts per link over time with different parameters in MC-DFP. Selected values  of parameters $(\rho_1,\rho_2,\eta_1,\eta_2)$ are  for (1) Top-Left: (0.5, 1 , 0.1, 0.4),  (2) Top-Right: (0.1, 0.2, 0.1 ,0.4), (3) Bottom-Left: (0.5, 1, 0.2 ,1.5), (4) Bottom-Right: (0.1, 0.2, 0.2 ,1.5). For each set of parameters, we show average communication attempt per link over 20 runs of Scenario 1 with speed $\alpha=0.1$. Percentage values in red in each figure show the success rate of NE convergence. The case with large fading constants combined with small threshold constants (Top-Left) is both effective and efficient. }\vspace{0pt}
	\label{fig_com_atp}
\end{figure}

We analyze the effects of fading memory constants $\rho_1$ and $\rho_2$, and threshold constants $\eta_1$ and $\eta_2$ in MC-DFP for Scenario 1. We consider large $(\rho_1,\rho_2)=(0.5, 1)$ and small $(\rho_1,\rho_2)=(0.1, 0.2)$ fading constant values along with large $(\eta_1,\eta_2)=(0.2, 1.5)$ and small $(\eta_1,\eta_2)=(0.1, 0.4)$ communication threshold constants. As fading memory constants take large values, robots dismiss past information faster. As threshold constants take small values, robots are less likely to cut communication as per \eqref{eq_w}. We observe that as threshold constants increase, the likelihood of successful convergence to NE drops significantly (compare percentage values in red in Fig.~\ref{fig_com_atp} Top and Bottom). Moreover, if threshold constants are low enough, then it is better to have high fading constants in terms of saving communication energy (compare Fig.~\ref{fig_com_atp} Top-Left and Top-Right). However, if threshold constants are high, then it is better to have small fading constants so that communication is not cut very early to prohibit convergence to NE (compare Fig.~\ref{fig_com_atp} Bottom-Left and Bottom-Right). Overall, small communication threshold values combined with high fading constants guarantee convergence while reducing communication attempts by a three-fold compared to DFP.

\subsection{Experiments}

% We also experiment with MC-DFP under real-life experiments. For this purpose, we utilized Robotarium project \cite{wilson2020robotarium}.
We use GRITSBot X model mobile wheeled robots, each 11 cm wide, 10 cm long, and 7 cm tall, made available by the Robotarium project \cite{wilson2020robotarium}.
% Robotarium project provides remote usage of the GRITSBot X model mobile wheeled robots, whose sizes are 11 cm wide, 10 cm long, and 7 cm tall.
The robots operate in a 3.2 m x 2 m area with maximum speed of 20 cm/s linearly and maximum rotational speed of about 3.6 rad/s. For more technical details, see \cite{wilson2020robotarium}.

We coded the implementation in MATLAB and, use the same set of parameter values used in Fig. \ref{fig_com_atp} Top-Left, i.e., $(\rho_1= 0.5, \rho_2=1 ,\eta_1= 0.1, \eta_2=0.4)$. Similarly, we select the inertia probability $\epsilon_{inertia}$ and the channel fading constant $r$ as $0.1$ and $0.60$, respectively. We let $\Delta_{ij}$ in \eqref{eq_w} be equal to 10 as before. The random communication model is simulated within MATLAB. 

We consider randomly assigned initial robot and target positions for a team of size $N=5$, $N=10$ and $N=18$ (maximum number allowed in Robotarium). The links for the experiments for each case can be found as below,
 
 \begin{enumerate}
     \item \url{https://youtu.be/wuJLI6CNGgU} ($5$ robots),
     \item \url{https://youtu.be/KyoPGJNql2c} ($10$ robots),
     \item \url{https://youtu.be/PSn_osWDGXA} ($18$ robots).
 \end{enumerate}

We observe that robots are able cover a sequence of randomly assigned targets for all the scenarios.  

There are several differences in the movements of robots between the simulations and the experiments. Due to linear and rotational speed limitations, robots are not always able to turn and go through the directions assigned by the algorithm. Moreover, local collision avoidance protocols, that prohibit robots getting too close, limit the mobility decisions. We observe that these differences do not affect the overall performance of MC-DFP. Robots successfully reach one-to-one assignment with targets and cover targets physically.

%However, while in a few cases are successfully  accomplished, i.e targets are covered, experiments are terminated earlier since robots attempt to go outside of the test area. The reason is that as the number of robots increases, the test area becomes too small for robots such that they cannot move to avoid collision inside the test area

%%%%%%%%%%%%%%%%%%%%%%%%%%%%%%%%%%%%%%%%%%%%%%%%%%%%%%%%%%%%%%%%%%%%%%%%%%%
%%%   S E C T I O N %%%%%%%%%%%%%%%%%%%%%%%%%%%%%%%%%%%%%
%%%%%%%%%%%%%%%%%%%%%%%%%%%%%%%%%%%%%%%%%%%%%%%%%%%%%%%%%%%%%%%%%%%%%%%%%%%
%

%%%   S E C T I O N %%%%%%%%%%%%%%%%%%%%%%%%%%%%%%%%%%%%%
%%%%%%%%%%%%%%%%%%%%%%%%%%%%%%%%%%%%%%%%%%%%%%%%%%%%%%%%%%%%%%%%%%%%%%%%%%%
%

%%%%%%%%%%%%%%%%%%%%%%%%%%%%%%%%%%%%%%%%%%%%%%%%%%%%%%%%%%%%%%%%%%%%%%%%%%%
%%%   S E C T I O N %%%%%%%%%%%%%%%%%%%%%%%%%%%%%%%%%%%%%
%%%%%%%%%%%%%%%%%%%%%%%%%%%%%%%%%%%%%%%%%%%%%%%%%%%%%%%%%%%%%%%%%%%%%%%%%%%
%
\section{Conclusion}
We proposed decentralized mobility and communication protocols for a team of robots solving a target assignment problem by best responding to the intended target selection of other robots. Each robot learns about others' intended selections by keeping track of others' frequency of past actions. For keeping such estimates, robots need to be able to transmit their empirical frequencies to each other over a wireless network subject to path loss and fading. The proposed communication protocol relies on metrics that measure novelty of information and information need of the robots to decide whether to transmit or not and how to allocate available communication resources. Moreover, robots may alter their mobility to overcome fading in communication depending on their assessment of the need to communicate certain robots. We stated sufficient conditions for convergence to a NE, and presented numerical and {experimental} results that demonstrated the benefits of the proposed learning-aware voluntary communication and the communication-aware mobility protocols on reducing communication need while retaining convergence guarantees.

\appendix

\input{appendix.tex}

\bibliographystyle{IEEEtran}
\bibliography{bibliography}

\end{document}

%% file: my_sections.tex
\usepackage{needspace}

\newcommand{\myparagraph}[1]{\needspace{1\baselineskip}\medskip\noindent {\it #1.}}

%% file: introduction.tex
% !TEX root = dfp_com_aware.tex

{With advances in robotics and wireless communication, autonomous systems are deployed in many different areas ranging from unmanned aerial vehicles (UAV) \cite{dutta2017decentralized} to watercraft systems\cite{hung2021cooperative,hu2021distributed} and self-driving cars \cite{carron2019scalable}.  In practice, typical goals of multi-agent robot teams can be search, rescue, and patrolling  missions. Accomplishment of these goals includes, but not limited to, scanning and covering physical locations in hazardous environments, where communication abilities are limited---see \cite{dorri2018multi,shi2020survey,rizk2018decision} for more current applications of autonomous systems. } 
In such team missions, autonomous robots are put together to collaboratively achieve a common goal utilizing wireless communication and their physical abilities. Collaboration entails each team member gathering data and resolving differences with others efficiently via rapid communication to produce a joint action profile.  Here, we posit that communication and mobility capabilities need to be managed by the team members based upon the occurrence of a need for additional information in order to maximize team performance. 

 In particular, we consider a team of robots tasked with covering a given set of targets. 
 Target assignment problems are  combinatorial optimization problems seeking assignments between resources (robots) and targets to maximize utitilies or minimize costs---see \cite{Murphey2000,conforti2014integer} for detailed surveys of centralized approaches on target assignment problems. Among the multi-agent approaches to solving the target assignment problem are auction-based algorithms \cite{choi2009consensus,luo2014provably,otte2020auctions}, dynamic partitioning and coalition formation \cite{liu2012large,mazdin2021distributed}, temporal-logic based approaches \cite{lindemann2019coupled}, and Voronoi-partitioning based control schemes \cite{abbasi2017new}. 
 
%  \blue{Different approaches on multi-robot setting have also also discussed.
%  employ auction-based algorithms in distributed systems.  \cite{liu2012large,mazdin2021distributed} are based on dynamic partitioning and coalition formation in both centralized and decentralized settings. \cite{lindemann2019coupled} utilizes temporal-logic to provide discrete tasks for heterogeneous robotic teams.\cite{abbasi2017new} treats area coverage of mobile sensor networks as a target-assignment problem and, propose Voronoi-partitioning based control schemes. }

  In this paper, robots have limited communication resources per decision epoch, and communication is subject to failures due to path-loss and fading. Figure \ref{fig_draw} shows an example of a team of three robots that wants to cover three targets. Here we model the target assignment problem as a game played among team members whose pure Nash equilibria correspond to robots covering all targets \cite{arslan2007autonomous} (Section \ref{sec_target_game}). 
Given the setup, along the lines of the aforementioned vision for team collaboration, we propose a decentralized game-theoretic learning algorithm in which agents learn to cover the targets as a team by making {\it learning-aware communication}, and {\it communication-aware mobility} decisions.

In particular, we generalize a decentralized form of the fictitious play (FP) algorithm, so that it is suitable for realistic communication and mobility settings, and is able to manage limited communication resources (Section \ref{sec_DECENTRALIZED}). The proposed algorithm has three main parts that operate in tandem: a) {\it Fictitious play}: agents keep estimates of the intended target selections of other agents to select best available targets; b) {\it Intermittent and voluntary communication:} agents use their current estimates to make voluntary communication attempts with other agents; c) {\it Communication-aware mobility:} agents take movement actions considering the trade-off between covering their selected targets in a given time and increasing chance of successful communication.

FP algorithm is a best-response type distributed game-theoretic learning algorithm \cite{Shamma_Arslan_2005,Swenson_et_al_2014,salehisadaghiani2019distributed,parise2019distributed}. In the classic FP algorithm \cite{monderer1996fictitious}, an agent takes an action that maximizes its expected utility assuming other agents select their actions randomly from a stationary distribution. In FP, agents assume this stationary distribution is given by the past empirical frequency of past actions. FP is not a decentralized algorithm, since agents need to observe past actions of everyone to be able to form these distributions, and compute their utility expectations. In the decentralized fictitious play (DFP) algorithm, agents form estimates on empirical frequencies of other agents' actions by averaging the estimates of their neighbors received over a communication network. The fast convergence rate of averaging updates guarantee convergence of DFP algorithm for potential games, i.e., games with payoffs that admit potential functions \cite{Swenson_et_al_2014,eksin2017distributed,swenson2018distributed}. Here, the generalization of the DFP algorithm allows for communication failures, and intermittent and voluntary communication attempts.

{\it Learning-aware communication} refers to agents assessing novelty of their information and the information need of other agents who are potential receivers of information. Indeed, if an agent has little new information to share, the individual agent can save energy by abstaining from communication without hampering team performance. Based on this premise, recent studies in distributed optimization  \cite{chen2021ordering,chen2018lag,liu2019communication} propose local threshold based communication protocols that rely on novelty of information measured by, e.g., change in local gradient. These communication protocols are also referred to as censoring since a sender agent self-censors if it deems its information as stale. 
Along the same lines, here we consider a communication protocol that provides full autonomy to agents in deciding whether to communicate or not based on changes in their estimates of empirical frequencies. Our protocol departs from the past approaches by the feature that agents also determine who to communicate with by assessing information need of other agents, in addition to the novelty of local information. Specifically, agents keep an estimate of the similarity between each others' estimates of empirical frequencies. If agent $i$'s assesses that agent $j$ may have similar estimates due to past interactions, it may choose to not transmit its new estimate to agent $j$. Moreover, each agent  allocates their communication resources based on a statistic that measures the similarity of target selections. That is, if an agent is more likely to select the same target with another agent, then it is more urgent for these two agents to coordinate their selections. These features in which agents determine who to communicate with and allocate communication resources based on urgency of information exchange makes the communication protocol learning-aware. 

{\it Communication-aware mobility} refers to agents determining their heading directions not just based on their target selections, but also based on their need to communicate in the presence of fading. In the target assignment game, if robots move toward their selected targets, they may quickly lose connectivity due to growing distances between them. This may lead to certain robots committing early to their target selections without spending the time needed to coordinate their actions with all the other agents. Since the team would need to resolve agent-target allocations eventually, this mobility would be highly inefficient as some agents may take many steps toward their selected targets only to change their selections. In addressing some of these issues, recent works in mobility and communication control in autonomous teams propose mobility decisions that mind communication \cite{fink2011robust,stephan2017concurrent,kantaros2016distributed,yan2014go}---also see \cite{muralidharan2021communication} for a survey on the relation of different communication setups and mobility. However, in these studies network connectivity is treated as a constraint to be satisfied by the team. Ensuring connectivity as mobile robots move to reach their selected targets can significantly hamper team performance, and cause the target assignment problem to be infeasible since some targets may never be covered to remain connected. Recent studies on mobile robotic teams account for intermittent communication for distributed state estimation problems \cite{kantaros2019temporal,khodayi2019distributed}. Along similar lines, we incorporate information exchange needs of agents and fading effects into their mobility decisions. Our goal is to reduce total effort spent by the team by increasing the chances of successful communication attempts. 

We analyze the convergence properties of the DFP algorithm equipped with learning-aware communication and communication-aware mobility protocols in the target assignment game (Section \ref{sec_con}). First we show that the target assignment game is a best-response potential game, i.e., there exists a potential function such that the best action for an agent is the same regardless of whether it optimizes its utility function or the potential function.  Then, we show that the DFP algorithm converges to a pure Nash equilibrium of the target assignment game given appropriate choices of the threshold parameters in the learning-aware communication protocol. Given the convergence to a pure Nash equilibrium in finite time, we are able to show that all targets are eventually covered by the team (Theorem \ref{cor_physical}).  Numerical {simulations} demonstrate the reduction in the number of communication attempts due to learning-aware communication, and the increased likelihood of finishing the given task by the final time due to communication-aware mobility. We demonstrate the practical applicability, the effectiveness and the scalability of the decentralized decision-making scheme in experiments with a team of mobile-wheeled robots (Section \ref{sec_numeric}).%\begin{footnote}{We considered similar communication schemes for the DFP algorithm in  \cite{aydin2020communication,aydin2020decentralizedcdc,aydin2021decentralized}. In \cite{aydin2020communication}, we propose a self-censoring protocol based only on the novelty of information for the target assignment game. The  communication protocol in this paper extends it to consider the value of local information for the receiving agents. In addition, we propose a novel communication aware mobility protocol where agents anticipate the role of their future positions in the chance of communication. In \cite{aydin2020decentralizedcdc,aydin2021decentralized}, we provide a theoretical framework for the convergence of the communication-censored DFP algorithms in weakly-acyclic games by providing a sufficient learning condition that needs to be satisfied by a communication protocol. In this paper, we propose communication protocols particular to the target-assignment game and rely on the theoretical framework developed in \cite{aydin2021decentralized} to prove convergence. Our contribution in this paper is to put forth efficient local communication protocols for autonomous teams aiming to solve the target assignment game. %We show the scalability of the proposed protocols and the reduction in communication attempts via simulations and experiments.

%% file: appendix.tex
\subsection{Proof of Lemma \ref{lem_cond}} \label{ap_pr_2}

 Let's define the following events $E_1$ and $E_2$ in order to show Condition \ref{cond_prob} holds,
 \begin{align}
     E_1(t)=\{&||a_j(t+T)-f_j(t+T)|| \le \xi/2\} \\
    E_2(t)=\{&||f_j(t+T)-f^{i}_{j}(t+T)|| \le \xi/2\}
 \end{align}
By the triangle inequality, we have
\begin{align}
    \mathbb{P} (||a_j(t+T)-f^i_j(t+T)|| \le {\xi}| {\ccalH(t))} &\ge  \nonumber\\
    &\mathbb{P} (E_1(t),E_2(t) ) | {\ccalH(t))}.
\end{align}
Hence, it is enough to show that the intersection of the events $E_1$ and $E_2$ has positive probability to assure Condition \ref{cond_prob}. Given the repetition of actions by an agent $j$, for any fading rate $\rho_1 \in (0,1]$, there exists a finitely long enough  $\hat{T}$ as the lower bound on agent $j$ repeating the same action $a_j\in \ccalA$ such that the following holds,
\begin{align}
    \mathbb{P} (E_1(t) ) | {\ccalH(t))} = 1.
\end{align}
Observe that the model \eqref{eq_rout_2} always admits optimal solutions $\beta^*_{ij}(t)>b$ where $b >0$, as long as the weights $w_{ij}(t) >0$. Combined with Assumption \ref{as_dist}, it holds,
\begin{equation}
    \beta^*_{ij}(t)\, e^{-r ||x_{i}(t)-x_{j}(t)||^2_2} \ge \epsilon_{com}= b e^{(-rD^2)} >0,
\end{equation}
when $w_{ij}(t) >0$ with small enough $0\le\eta_1<\xi/2$, and $0\le \eta_2\le \xi/2$ by the definition \eqref{eq_w}. Further note that Assumption \ref{as_ack} lets agent $j$ to acknowledge its successful communication so that agent $j$ can also locally compute agent $i$'s information on itself $f^{i}_{j}(t)$ for any time t. For any $\rho_2 \in (0,1]$, during consecutive repetition of the same action $a_j=\bbe_k$ from time $t$ to $t+T$, starting at time $t+T_1$ where $T=T_1+T_2$ and $T_1 <T$, agent $j$ needs to send $f_j$ for $T_2$ times ending at $t+T$. This provides that, the event $E_2(t)$ has a positive probability, 
\begin{align}
    \mathbb{P} (E_2(t) ) | {\ccalH(t))} \ge \epsilon_{com}^{T_2}.
\end{align}
Thus, there exists a positive bound $\hat{\epsilon}$ on the probability of the given event in Condition \ref{cond_prob},
\begin{align}
    &\mathbb{P} (||a_j(t+T)-f^i_j(t+T)|| \le {\xi}| \ccalH(t)) \ge \nonumber\\
    &\mathbb{P}(E_1(t),E_2(t)| \ccalH(t)) \ge \epsilon_{com}^{T_2} =\hat{\epsilon}>0.
\end{align}